\documentclass{article}
\usepackage{spconf,amsmath,graphicx}

\usepackage{enumitem}
\setlist{nosep, leftmargin=14pt}

\usepackage{mwe}

\title{
Investigating the Benefit of Transfer Learning from the Locus Coeruleus for Substantia Nigra Pars Compacta Segmentation}

\name{Tayebeh Bahador\textsuperscript{2},
Max Dünnwald\textsuperscript{1,3},
Matthew J. Betts\textsuperscript{4,5,8},
Emrah Düzel\textsuperscript{4,5,7,8},
Steffen Oeltze-Jafra\textsuperscript{6,1}}

\address{\small
\begin{tabular}{@{}c@{}}
$^{1}$Department of Neurology, Faculty of Medicine, Otto von Guericke University (OVGU)\\
$^{2}$Faculty of Electrical Engineering and Information Technology, Institute for Medical Engineering, OVGU\\
$^{3}$Faculty of Computer Science, OVGU\\
$^{4}$German Center for Neurodegenerative Diseases (DZNE), Magdeburg, Germany\\
$^{5}$Institute of Cognitive Neurology and Dementia Research, OVGU\\
$^{6}$Peter L. Reichertz Institute for Medical Informatics, Hannover Medical School\\
$^{7}$Institute of Cognitive Neuroscience, University College London, London, UK\\
$^{8}$Center for Behavioral Brain Sciences (CBBS), OVGU, Magdeburg, Germany\\
tayebeh.bahador71@gmail.com
\end{tabular}}

\begin{document}

\maketitle
{\renewcommand{\thefootnote}{}\footnotetext{This work has been
submitted to the IEEE International Symposium on Biomedical Imaging
(ISBI) 2027 for possible publication. Copyright may be transferred without notice, after which this version may no longer be accessible.}}

\begin{abstract}
The substantia nigra pars compacta (SNpc) is a small midbrain nucleus that degenerates early in Parkinson's disease (PD). T1-weighted FLASH MRI enables its visualization and may support early detection of neurodegeneration. However, manual segmentation is difficult and time-consuming, and deep learning models often suffer from the limited availability of labeled data. Transfer learning (TL) may help by adapting pre-trained models to small datasets.

We investigate whether TL from a model trained on the locus coeruleus (LC), a nucleus with a similar appearance in this sequence, improves SNpc segmentation. Several TL strategies were evaluated using a 3D U-Net. With as few as 5--14 labeled subjects, TL and training from scratch achieved similar performance, reaching a Dice similarity coefficient of up to 0.77, close to reported inter-rater agreement. No TL strategy differed significantly from training from scratch, suggesting that initialization from the LC offers no significant benefit for SNpc segmentation.

\end{abstract}
\section{Introduction}
\label{sec:intro}
Parkinson’s disease (PD) is becoming more common as the population ages \cite{REEVE201419}. In PD, the substantia nigra pars compacta (SNpc), a small midbrain nucleus, shows severe depletion of neuromelanin, the dark pigment of its dopaminergic neurons \cite{zecca2002absolute}. The locus coeruleus (LC), a small brainstem nucleus with similarly pigmented neurons, degenerates in PD as well. Since current treatments only relieve symptoms, early-stage imaging biomarkers of these nuclei may enable earlier diagnosis and intervention \cite{betts2019locus}.
Both nuclei can be visualized with high-resolution T1-weighted MRI, where they appear hyperintense \cite{betts2019locus, sasaki2006neuromelanin}. Quantifying them requires accurate segmentation, but manual segmentation of these tiny structures is time-consuming and variable, with inter-rater Dice scores of about 0.83 for the SNpc \cite{le2019convolutional}. Therefore, automated segmentation methods are needed.
\newline Deep learning, in particular U-Net-based architectures \cite{ronneberger2015unet, cicek20163d} and frameworks such as nnU-Net \cite{isensee2021nnu}, has become the standard for medical image segmentation. More recently, foundation models such as SAM and MedSAM \cite{kirillov2023segment, ma2024segment} have enabled promptable segmentation. However, they still require adaptation to small, low-contrast structures, and labeled data remains the bottleneck. U-Net-based networks have been successfully applied to segmentation of the substantia nigra \cite{krupivcka2019automatic} and the SNpc \cite{le2019convolutional}, reaching Dice scores comparable to inter-rater agreement, as well as to segmentation of LC \cite{dunnwald2020automated, dunnwald2021fully}. Building on the 3D U-Net established for the LC \cite{dunnwald2020automated, dunnwald2021fully}, we compare fine-tuning and layer-freezing strategies for transferring this model from LC to SNpc segmentation across decreasing training set sizes.

In transfer learning, the network is initialized with weights of another model, which was trained on another task, and may provide useful prior knowledge when target labels are scarce \cite{tajbakhsh2016convolutional}. 
Given that both nuclei have a similar appearance in this sequence, we hypothesize that a model trained to segment the LC may provide a useful initialization for SNpc segmentation, particularly when only a few labeled subjects are available. As both structures also change with normal aging \cite{zecca2002absolute, betts2017vivo}, we study this hypothesis in healthy young and older adults as a first step before patient cohorts.
To the best of our knowledge, no prior work has examined whether a model trained on LC can be transferred to SNpc segmentation, nor how performance of TL depends on the number of available SNpc training subjects.
Therefore, the objectives of this work are:
\begin{enumerate}
    \item Comparison of different TL setups for SNpc segmentation based on LC initialization.
    \item Comparison of LC-based TL with SNpc segmentation from scratch.
    \item Analysis of different SNpc training set sizes to determine whether
      TL outperforms training from scratch at any size, and how few
      labeled subjects suffice.
    
\end{enumerate} 

\section{Materials and Methods}
\label{sec:methods}

\begin{figure}[t]
    \centering
    \includegraphics[width=0.8\linewidth]{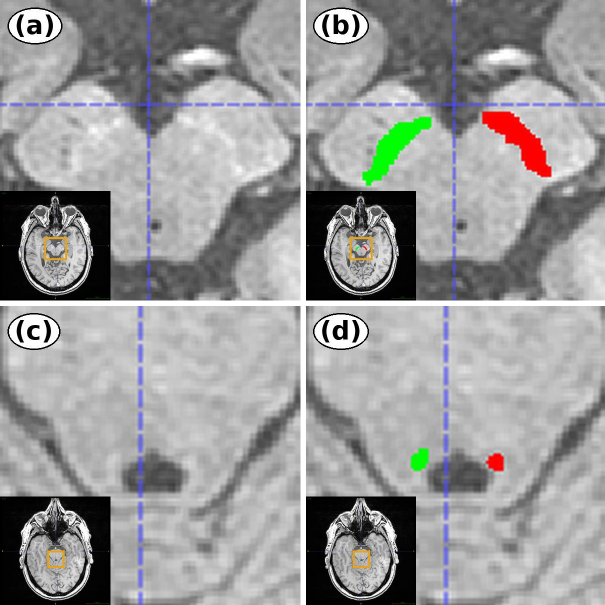}
    \caption{Axial T1-weighted FLASH slices: (a) SNpc without and (b) with
    manual segmentation; (c) LC without and (d) with manual segmentation.}
    \label{fig:Input_groundtruth}
\end{figure}

We used T1-weighted FLASH images of 81 healthy adults (22--30 and 61--80 years) from the cohort described in \cite{betts2017vivo}, acquired at 3T with 0.75\,mm isotropic resolution. Images were upsampled to 0.375\,mm isotropic resolution by sinc interpolation. LC and SNpc were manually segmented by a trained expert (Figure.~\ref{fig:Input_groundtruth}).

\begin{figure}[t]
  \centering
  \includegraphics[width=1.0\columnwidth]{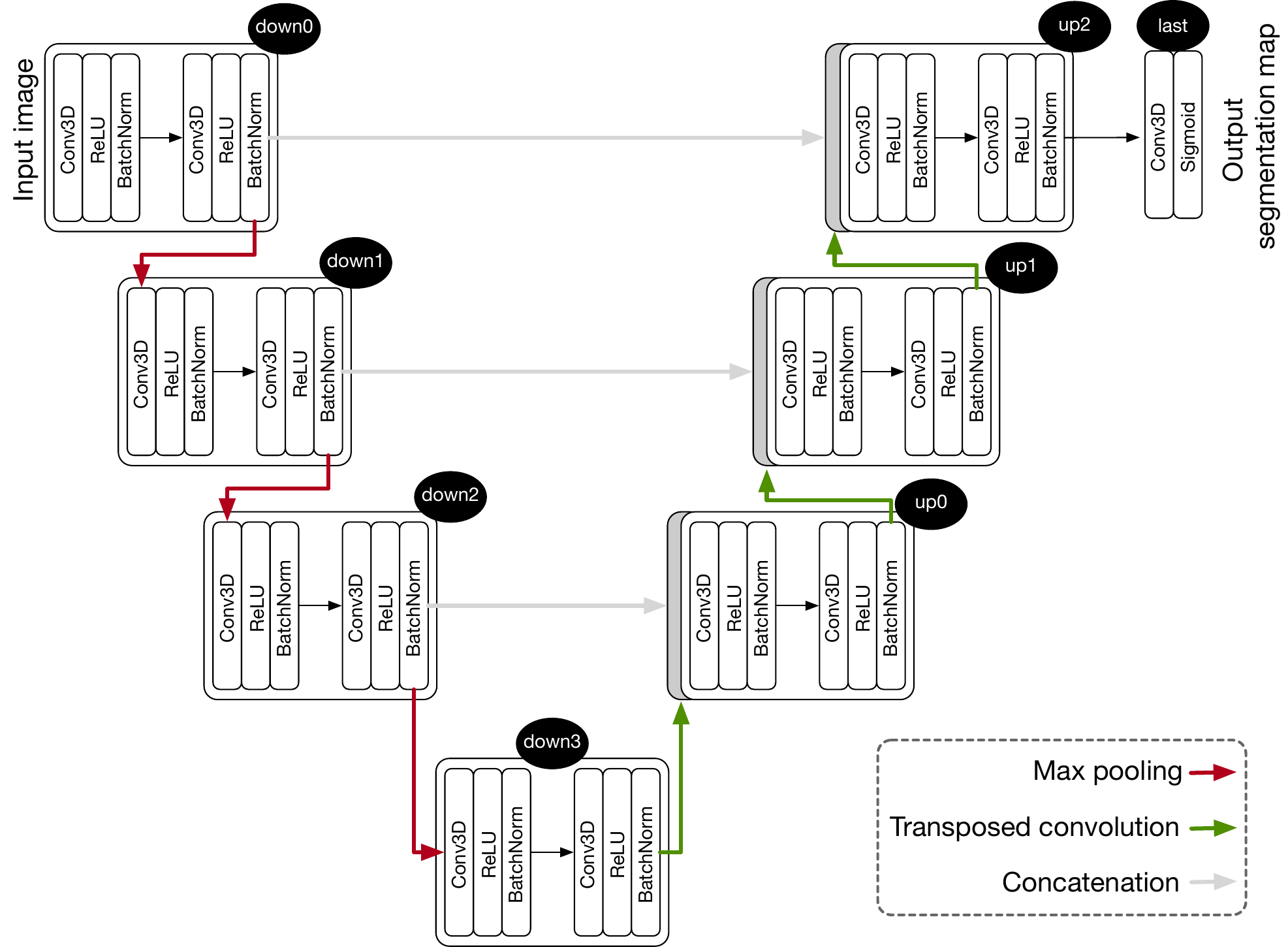}
  \caption{3D U-Net architecture used for both LC and SNpc segmentation. Block labels (e.g., down1, up2) are used to name the transfer learning strategies.}
  \label{fig:my_network}
\end{figure}

We adopt the 3D U-Net architecture of Dünnwald et al. \cite{dunnwald2020automated, dunnwald2021fully}, which was developed for LC segmentation on the same T1-weighted FLASH data. 
The network (Figure.~\ref{fig:my_network}) consists of three down-sampling blocks, a bottleneck block, and three up-sampling blocks, followed by a final convolutional layer with a sigmoid function. Each block contains two $3 \times 3 \times 3$ convolutional layers, each followed by ReLU activation and batch normalization \cite{ioffe2015batch}. $2 \times 2 \times 2$ max pooling and transposed convolutions are used for down- and up-sampling, and encoder feature maps are concatenated with the corresponding decoder features. Padding keeps input and output sizes identical.

\noindent Models were trained with the Dice loss (DSC)
\begin{equation}
  \mathcal{L} = 1 - \frac{2\sum_i p_i g_i + 1}{\sum_i p_i + \sum_i g_i + 1},
\end{equation}
where $p_i \in [0,1]$ is the predicted label and $g_i \in \{0,1\}$ the ground-truth label of voxel $i$. The constant 1 avoids division by zero. DSC was also used for evaluation.

Following \cite{dunnwald2020automated, dunnwald2021fully}, we used Adam \cite{kingma2014adam} with a learning rate of $10^{-3}$. Each mini-batch contained one randomly sampled $64^3$ patch per training subject, containing the target structure with a probability of 0.5.

We first trained a model from scratch for LC segmentation on all 81 subjects for 2,500 epochs. Its final weights were used to initialize SNpc segmentation models.

Two transfer learning strategies were evaluated. In the freeze strategy, separate models were trained in which all blocks from a given block to the output were fine-tuned with a learning rate of $10^{-3}$, following \cite{dunnwald2020automated}, while the preceding blocks were kept frozen (learning rate 0). We denote each model by its first fine-tuned block, from last (fine-tuning the final layer only) through up2 (fine-tuning the final layer and up2 block), up1 (fine-tuning the final layer, up2 and up1 blocks), up0, down3, down2 and down1 to whole (fine-tuning the entire network; Figure.~\ref{fig:my_network}). The decay strategy used the same configurations, but the preceding blocks were trained with a tenfold smaller learning rate ($10^{-4}$) rather than zero.

To select promising strategies, all TL setups and the training from scratch were first trained on 81, 62, 42, and 22 subjects for 2,500 epochs, each with a random 27\% test split. The TL strategies freeze-last (fine-tuning the last layer and freezing the rest), freeze-up2 (fine-tuning the last layer and up2 and freezing the rest), freeze-up1, and freeze-up0 consistently performed poorly. Therefore, these strategies were excluded, and the remaining strategies were chosen as the best for fine-tuning with a larger dataset. 

In the main experiments, the remaining more promising TL strategies and the training from scratch as a baseline were trained for 10,000 epochs on 14, 7, and 5 subjects, obtained as nested subsets of the 81-subject training set. For these training runs, from 81 subjects, 59 subjects were chosen as the training set, and 22 subjects were held out as a fixed test set. The 14 subjects were chosen from the 59-subject training set, 7 subjects from the 14-subject training set, and 5 subjects from the 7-subject training set. Therefore, all models across the 14-, 7-, and 5-sample-sized training sets were evaluated on the same test subjects, which none of the SNpc models had seen during training. Each subset was split into training and validation subjects by 5-fold cross-validation. The checkpoint with the lowest validation loss was retained.

For testing, a $128^3$ patch centered on the target structure was segmented, the sigmoid output was binarized and thresholded at 0.5, and connected components of 50 voxels or fewer were discarded. 

Segmentation accuracy was measured by the DSC on the combined left and right SNpc, averaged over the 22 test subjects for each fold, and reported as mean $\pm$ SD across the 5 folds. Differences between each TL strategy and training from scratch were assessed with Welch's $t$-test on fold-level DSC ($n=5$), with Holm correction across all 36 comparisons (12 strategies across 3 training set sizes). A~\mbox{$p$-value $<0.05$} was considered significant.

\section{Results}
\label{sec:results}

\begin{figure}[t]
  \centering
  \includegraphics[width=\columnwidth]{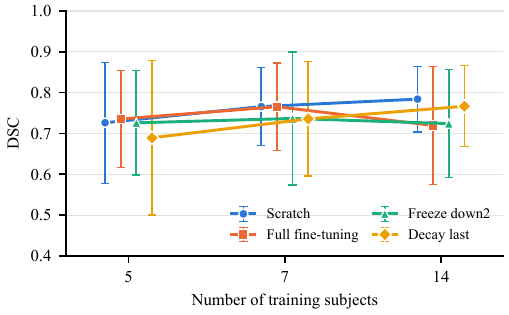}
  \caption{Mean $\pm$ SD DSC of training from scratch and selected TL strategies. All strategies are listed in Table~\ref{tab:results}.}
  \label{fig:dsc_size}
\end{figure}

Figure~\ref{fig:dsc_size} and Table~\ref{tab:results} summarize the DSC of all strategies on the fixed test set. Training from scratch achieved a mean DSC of 0.78 with 14 subjects, decreasing only moderately to 0.73 with 5 subjects. The best transfer learning configurations reached accuracy comparable to training from scratch (e.g., 0.77 with 14 subjects for Decay last, 0.77 with 7 and 0.74 with 5 subjects for full fine-tuning), and no strategy differed significantly from training from scratch at any size (Welch's $t$-test, all uncorrected $p \geq 0.19$; all Holm-corrected $p = 1.00$). With 5 subjects, decay down3 and decay down2 yielded the lowest mean DSC (0.56) and the largest SD across folds (0.31--0.32), indicating unstable training. Whole (freeze) and whole (decay), which both fine-tune the entire network, differed by up to 0.06 in mean DSC.

\begin{table}[t]
\centering
\caption{Mean $\pm$ SD DSC per strategy and training set size.}
\vspace{2pt}
\label{tab:results}
\small
\setlength{\tabcolsep}{4pt}
\begin{tabular}{lccc}
\hline
Strategy & 14 subjects & 7 subjects & 5 subjects \\
\hline
Scratch        & 0.78 $\pm$ 0.08 & 0.77 $\pm$ 0.10 & 0.73 $\pm$ 0.15 \\
\hline
Decay last     & 0.77 $\pm$ 0.10 & 0.74 $\pm$ 0.14 & 0.69 $\pm$ 0.19 \\
Decay up2      & 0.76 $\pm$ 0.12 & 0.66 $\pm$ 0.20 & 0.58 $\pm$ 0.22 \\
Decay up1      & 0.65 $\pm$ 0.21 & 0.66 $\pm$ 0.18 & 0.65 $\pm$ 0.21 \\
Decay up0      & 0.68 $\pm$ 0.17 & 0.72 $\pm$ 0.16 & 0.65 $\pm$ 0.20 \\
Decay down3    & 0.71 $\pm$ 0.15 & 0.69 $\pm$ 0.17 & 0.56 $\pm$ 0.31 \\
Decay down2    & 0.72 $\pm$ 0.14 & 0.74 $\pm$ 0.15 & 0.56 $\pm$ 0.32 \\
Decay down1    & 0.70 $\pm$ 0.17 & 0.68 $\pm$ 0.19 & 0.69 $\pm$ 0.17 \\
Whole (decay)  & 0.66 $\pm$ 0.17 & 0.77 $\pm$ 0.10 & 0.69 $\pm$ 0.16 \\
\hline
Freeze down3   & 0.69 $\pm$ 0.15 & 0.72 $\pm$ 0.16 & 0.65 $\pm$ 0.17 \\
Freeze down2   & 0.72 $\pm$ 0.13 & 0.74 $\pm$ 0.16 & 0.73 $\pm$ 0.13 \\
Freeze down1   & 0.71 $\pm$ 0.16 & 0.73 $\pm$ 0.16 & 0.71 $\pm$ 0.17 \\
Whole (freeze) & 0.72 $\pm$ 0.14 & 0.77 $\pm$ 0.11 & 0.74 $\pm$ 0.12 \\
\hline
\end{tabular}
\end{table}

\section{Discussion and Conclusion}
\label{sec:discussion}
Initializing SNpc segmentation with LC-trained weights did not yield a measurable benefit over training from scratch, even with only 5 labeled training subjects. Our results further suggest that a 3D U-Net trained from scratch on few labeled subjects, only 5--14, reached a DSC of 0.73--0.78, approaching the inter-rater agreement of 0.83 reported for SNpc segmentation \cite{le2019convolutional}, which leaves only a narrow margin for improvement by transfer learning. A possible reason could be that the LC and SNpc differ considerably in size and shape despite their similar appearance.

Whole (freeze) and whole (decay), which both fine-tune the entire network with the same learning rate, used the same subjects and LC initialization in each fold but differed by up to 0.06 in the mean DSC. Since the random seed was set once for all sequentially trained configurations, the two runs of fine-tuning the whole network drew different random patches in a different order. The effect of this sampling variability is amplified with few training subjects. Each mini-batch contained only a few patches (one per subject), and the best checkpoint was selected on randomly sampled validation patches from very few subjects, so this sampling variability can strongly affect the final model. Given this variability, small differences between strategies may reflect this random variability rather than true effects. Therefore, the ranking of transfer learning strategies, including freezing versus learning rate decay, should be interpreted with caution.

This study has limitations. All data came from healthy adults of a single cohort and scanner. In patients, reduced SNpc contrast makes segmentation harder, so our findings need confirmation in patient cohorts. Second, the labels used in this study were from a single rater. So, the DSC reflects agreement with one expert, and makes the comparison with inter-rater agreement only indicative. Third, test patches centered on the target structure may make absolute DSC values slightly optimistic, but apply equally to all strategies. Finally, as LC and SNpc models used the same datasets, the LC source models may have seen images of SNpc test subjects (though never with SNpc labels), which could only favor transfer learning and thus does not weaken our conclusion.

In summary, transfer learning from the LC offered no significant advantage for SNpc segmentation, and training from scratch with as few as 5--14 labeled subjects yielded DSC values approaching the inter-rater agreement. Future work should validate these findings in patient cohorts and across scanners, and explore self-supervised pretraining as an alternative to transfer from a related structure.
Moreover, while we focused on the SNpc as the target structure because its dopaminergic loss is the hallmark of PD, TL in the reverse direction (SNpc to LC) would also be worth investigating, as it would move from a larger to a smaller structure, exploit the higher contrast-to-noise ratio of the SNpc, and probably could benefit from public datasets with SNpc labels for pretraining.

\section{Compliance with ethical standards}
\label{sec:ethics}

This study was performed retrospectively on data from a previously
published study \cite{betts2017vivo}, which received ethical approval
from the Faculty of Medicine of the University of Magdeburg. All participants gave written informed consent in accordance with the Declaration of Helsinki.

\section{Acknowledgments}
\label{sec:acknowledgments}
This work was conducted as part of a master's thesis at Otto von Guericke University. This research was supported by the federal state of Saxony-Anhalt, Germany (Project I88). M.\,J.\,B.\ and E.\,D.\ are supported by the Deutsche Forschungsgemeinschaft (DFG, German Research Foundation) -- Project-ID 425899994, Sonderforschungsbereich 1436 (SFB 1436); M.\,J.\,B.\ is additionally supported by the German Federal Ministry of Education and Research (BMBF, funding code 01ED2102B) under the aegis of the EU Joint Programme -- Neurodegenerative Disease Research (JPND). We thank the German Center for Neurodegenerative Diseases (DZNE) for providing the data. The authors have no relevant financial or non-financial interests to disclose. 

{\small
\bibliographystyle{IEEEbib}
\bibliography{refs}
}

\end{document}